\documentclass[12pt,tightenlines,eqsecnum,aps,amsmath,amssymb,nofootinbib,prd]{revtex4}

\usepackage{setspace}
\usepackage{amsmath,amssymb,amsfonts,amsthm,mathrsfs}
\usepackage{graphicx}
\usepackage{color,xcolor}

\begin{document}

\title{Loop quantum black hole extensions within the improved dynamics
}

\author{Rodolfo Gambini$^{1}$, Javier Olmedo$^{2}$, Jorge Pullin$^{3}$}
\affiliation {
  1. University of the Republic, Montevideo, Uruguay \\
  2. Departamento de F\'isica Te\'orica y del Cosmos, Universidad de Granada,  Granada-18071, Spain\\
  3. Department of Physics and Astronomy, Louisiana State University,
  Baton Rouge, LA 70803-4001, USA}

\begin{abstract}
We continue our investigation of an improved quantization scheme for spherically symmetric loop quantum gravity. We find that in the region where the black hole singularity appears in the classical theory, the quantum theory contains semi-classical states that approximate general relativity coupled to an effective anisotropic fluid.  The singularity is eliminated and the space-time can be continued into a white hole space-time. This is similar to previously considered scenarios based on a loop quantum gravity quantization.
\end{abstract}
\maketitle
\section{Introduction}

In a previous paper \cite{us-imp} we studied an improved quantization for spherically symmetric loop quantum gravity. Earlier work \cite{us} had considered a constant polymerization parameter, similarly to the ``$\mu_0$'' quantization scheme in loop quantum cosmology, whereas the improved quantization is similar to the ``$\bar{\mu}$'' quantization scheme \cite{ashtekarsingh}. Other approaches involving improved quantizations have also been explored in \cite{improved}. We observed that the singularity was removed, but we did not analyze in detail what happened to the space-time beyond the region where the singularity used to be. Here we complete that study. We find that in that region there exist semi-classical quantum states for which the theory behaves like a quantum version of general relativity coupled to an effective anisotropic fluid \cite{BH-fluid}
that violates the dominant energy condition. In the highest curvature region there is a space-like transition surface, something that was unnoticed in \cite{us-imp}. The space-time continues into a white hole geometry, like in Ref. \cite{aos}. However, in this work we consider a different regularization for the parametrized observable associated to the shift function. Its very definition requires the choice of a slicing and the new regularization avoids an undesirable dependence on it in the semiclassical limit.

The organization of this paper is as follows. In section II we discuss the physical sector of the quantum theory, focusing on semiclassical sectors. In section III we introduce a horizon penetrating slicing based on Painlev\'e--Gullstrand coordinates and show how it can be used to connect to a white hole space-time. We end with a discussion.

\section{Physical sector of the quantum theory}

The physical sector of the theory is obtained after combining Loop Quantum Gravity quantization techniques and the Dirac quantization programme for constrained theories. In summary, we start with a kinematical Hilbert space in the loop representation adapted to spherically symmetric spacetimes for the geometrical sector $[(K\varphi,E^\varphi),(K_x,E^x)]$ together with a standard representation for the spacetime mass and its conjugate momentum  $(M,P_M)$. A suitable basis of kinematical states is the one provided by spherical symmetric spin networks tensor product with the standard states for the matter sector in the mass representation. Then, we represent the scalar constraint as a well-defined operator in the kinemtical Hilbert space (for the diffeomorphism constraint we rather work with the related finite group of transformations mimicking the full theory). 

Following the construction of Ref. \cite{us-imp}, the physical sector of the theory is encoded in physical states (solutions to the scalar constraint) endowed with a suitable inner product and a set of physical observables. This is achieved, for instance, by applying group averaging techniques for both the quantum scalar constraint and the group of finite spatial diffeomorphisms (see also Refs. \cite{us,gowdy}). We focus our study to some of the simplest semi-classical states. Quantum states consist of spatial spin networks labeled by the ADM mass $M$ (a Dirac observable) and integer numbers that characterize the radii of spheres of symmetry associated with each vertex of the network $k_i$. The semi-classical states we are going to consider here are given by superpositions in the mass centered at $M_0$ and of width $\delta M_0$ and are therefore associated with a fixed discrete structure in space (see \cite{us-imp} for more details). They provide excellent approximations to the classical geometry in regions of {small curvature compared to Planck scale. Concretely, we consider the semi-classical states}
\begin{eqnarray}\nonumber
|\psi\rangle&=&\frac{1}{\delta M_0}\int dM e^{iM P_0/\hbar}\cos\left[\frac{\pi(M-M_0)}{2\delta M_0}\right]\Theta(M-M_0+\delta M_0)\Theta(M_0+\delta M_0-M)\\\label{eq:psi}
&&\times|M,k_S,\ldots,k_0,\ldots,k_{-S}\rangle
\end{eqnarray}
with $k_0<k_{j}$ for all $j\neq 0$, namely,  $j=-S,-S+1,\ldots,1,-1,\ldots,S$, where 
\begin{equation}
k_0={\rm Int}\left[\left(\frac{2 G M_0 \Delta}{4\pi \ell_{\rm Pl}^3}\right)^{2/3}\right],
\end{equation}
times the Planck length squared determines the smallest area of the 2-spheres in the  theory. This corresponds to the improved quantization, where $\Delta$ is the area gap. Besides,  we choose $M_0\gg m_{\rm Pl}$ and 
\begin{equation}
\delta M_0 \leq \frac{3}{2}\left(\frac{4\pi \ell_{\rm Pl}^3}{2G\Delta}\right)^{2/3}M_0^{1/3}.
\end{equation}
The states in Eq. \eqref{eq:psi} belong to a family of sharply peaked semiclassical states in the mass and with support on a concrete spin network (states with higher dispersion in the mass will require superpositions of different spin networks). This choice considerably simplifies the analysis of the effective geometries. As we discussed in our previous papers, the quantum theory has additional observables to the ones encountered in classical treatments \cite{kucharthiemann}
which are the ADM mass and the time at infinity. These emerge from the discrete nature of the spin network treatment  and are associated with the $k_i$'s, which in turn are associated with the value of the areas of the spheres of symmetry connected with each vertex of the spin network. One can also consider states that are a superposition of $M$'s. The analysis will remain the same as long as the states are peaked around a value of $M$.

In addition to physical states, the physical observables representing space-time metric components will be defined through suitable parametrized observables. They act as local operators on each vertex of the spin network. Furthermore, they involve point holonomies that are chosen to be compatible with the superselection sectors of the physical Hilbert space (see Ref. \cite{us-imp} for more details). {Some of the basic parametrized observables are 
\begin{align}\label{eq:hex}
  &\hat E^x(x_j)|M,\vec k\rangle=\hat O(z(x_j))|M,\vec k\rangle=\ell_{\rm Pl}^2 k_{j(x_j)}|M,\vec k\rangle
  ,\\\label{eq:hdex}
  & \hat M|M,\vec k\rangle=M|M,\vec k\rangle,
\end{align}
where $z(x)$ is a suitable gauge function that codifies the freedom in the choice of radial reparametrizations.
}

For the components of the space-time metric on stationary slicings we have, for instance, the lapse and shift,\footnote{In Ref. \cite{us-imp} for the shift we adopted the regularization $K_\varphi(x_j)\to \sin\left(2{\bar\rho}_j K_\varphi(x_j)\right)/2{\bar\rho}_j$, but it introduces an undesirable slicing dependence that is avoided with the present regularization. Besides, the representation that we adopt here for the square of the shift function as a parametrized observable is compatible with the superselection rules of the quantum numbers $\nu_j$ of the kinematical spin networks as it was discussed in Ref. \cite{us-imp}.}
\begin{equation}\label{eq:q-lapshi}
  \hat N^2(x_j) :=\frac{1}{4}\frac{([\hat E^x(x_j)]')^2}{(\hat E^\varphi(x_j))^2},\quad
  {[\hat N^x(x_j)]^2 ={\frac{\hat E^x(x_j)}{(\hat E^\varphi(x_j))^2}}\widehat{\frac{\sin^2\left({\bar\rho}_j K_\varphi(x_j)\right)}{{\bar\rho}^2_j}}},
\end{equation}
where
\begin{equation}
\label{eq:hephi}
(\hat E^\varphi(x_j))^2 = \frac{\left[(\hat E^x(x_j))'\right]^2/4}{1+\frac{{\sin^2\left(\bar\rho_j K_\varphi(x_j)\right)}}{\bar\rho_j^2} -\frac{2 G \hat M}{\sqrt{|\hat E^x(x_j)|}}},
\end{equation}
{where we polymerized $K_\varphi$} with $\bar \rho$ the polymerization parameter of the improved quantization,
\begin{equation}
\bar \rho = \frac{\Delta}{4\pi \hat E^x}.
\end{equation}
We choose the $k$'s in the one-dimensional spin network in our physical state and the gauge function $z(x)$ such that {
\begin{align}\label{eq:hex2}
  &\hat E^x(x_j)=\ell_{\rm Pl}^2{\rm Int}\left[\frac{x_j^2}{\ell_{\rm Pl}^2}\right],
  \\\label{eq:hdex2}
  & [\hat E^x(x_j)]'|M,\vec k\rangle=\frac{\ell_{\rm Pl}^2}{\delta x}\,{\rm Int}\left[\frac{(x_j+\delta x)^2-x_j^2}{\ell_{\rm Pl}^2}\right]|M,\vec k\rangle
  ,
\end{align}
}and
with $x_j=\delta x\,|j|+x_0$ and with $j\in\mathbb{Z}$, where 
\begin{equation}
x_0=\sqrt{{\rm Int}\left[\left(\frac{2 G M \Delta}{4\pi}\right)^{2/3}\right]}.
\end{equation}

Besides, we will choose $\delta x=\ell_{\rm Pl}$ as in the first paper \cite{us-imp}, although we will discuss the consequences of the limiting choices (for a uniform lattice) $\delta x=\frac{\ell_{\rm Pl}^2}{2x_0}$ and $\delta x=x_0$. The different spacings $\delta x$ just mentioned here correspond to different choices of states in the physical space, all of them lead to the same semiclassical behavior but differ in the deep quantum regime close to the singularity, as one would expect. The quantum regime is for the small values of $k_i$, where if one were to consider a superposition of states, small changes in $k_i$'s would lead to great fluctuations in the properties of the states.

Then, the metric components take the following form in terms of the previous operators
\begin{eqnarray}
\hat g_{tt}(x_j) &=& -\left(\hat N^2-\hat g_{xx}[\hat N^x]^2\right),\quad  \hat g_{tx}(x_j) = \hat g_{xx}{\sqrt{[\hat N^x]^2}},\nonumber\\
\hat g_{xx}(x_j) &=& \frac{(\hat E^\varphi)^2}{\hat E^x}, \quad g_{\theta\theta}(x_j)=\hat E^x,\quad g_{\phi\phi}(x_j)=\hat E^x\sin^2\theta.\nonumber
\end{eqnarray}

Let us restrict the study to the family of stationary slicings determined by the condition 
\begin{equation}\label{eq:slice}
\widehat{\sin^2\left(\bar\rho_j K_\varphi(x_j)\right)}=[\hat F(x_j)]^2
\end{equation}
where some specific choices of $F(x_j)$ will be studied below. However, any viable choice must be such that $F(x)$ is real and $F(x)\in[-1,1]$. Now, one can easily construct the operators corresponding to the components of the spacetime metric. They  are given by
\begin{eqnarray}
\hat g_{tt}(x_j) &=& -\left(1-\frac{\hat r_S}{\sqrt{\hat E^x}}\right),\quad  \hat g_{tx}(x_j) = -\sqrt{\frac{\pi}{\Delta}}\frac{\left\{\widehat{\left[E^x\right]'}\right\}{\sqrt{\hat F^2}}}{\sqrt{1-\frac{\hat r_S}{\sqrt{\hat E^x}}+\frac{4\pi \hat E^x \hat F^2}{\Delta}}},\nonumber\\
\hat g_{xx}(x_j) &=& \frac{\left\{\widehat{\left[E^x\right]'}\right\}^2}{4 \hat E^x\left(1-\frac{\hat r_S}{\sqrt{\hat E^x}}+\frac{4\pi \hat E^x \hat F^2}{\Delta}\right)}, \quad\hat g_{\theta\theta}(x_j)=\hat E^x,\quad \hat g_{\phi\phi}(x_j)=\hat E^x\sin^2\theta,\nonumber
\end{eqnarray}
with $\hat r_S = 2G\hat M$. The effective metric is defined as $g_{\mu\nu}=\langle \hat g_{\mu\nu}\rangle$, where the expectation value is computed on the extended physical state $|\psi\rangle$ we presented above. We will focus on the leading order corrections when the dispersion in the mass can be neglected. In this case, we can just remove the hats in the previous expression and denote this contribution by ${}^{(0)} g_{\mu\nu}(x_j)$. In addition, we will take a continuum limit that was discussed in our first paper. Namely, $x_j=\delta x\, |j|+x_0$ is replaced by $(|x|+x_0)$, with $x\in \mathbb{R}$ { and the integer part function ${\rm Int}[\cdot]$ will be dropped from all expressions. This continuum limit means} that the effective geometries bounce when they reach $x=0$. 

\section{Painlev\'e-Gullstrand coordinates: black hole to white hole transition}\label{sec:class}

We are interested in spatial slicings that are horizon penetrating and asymptotically flat. For instance, ingoing Painlev\'e-Gullstrand coordinates is one of the well-known choices that meet these requirements. Besides, the time coordinate follows the proper time of a free-falling observer. The slicing is defined by the condition  $\hat F(x_j) = \hat F_1(x_j)$ where
\begin{align}\label{eq:gf-f1}
\hat F_1(x_j)=\bar\rho\sqrt{\frac{\hat r_S}{\sqrt{\hat E^x}}}.
\end{align}
This choice is equivalent to a lapse operator $\hat N(x_j)=\hat I$. Besides, one can easily see that in the semiclassical limit $x_j\to x+x_0$ we have the function $F_1(x)<1$ for all $x\neq 0$, while $F_1(x=0)=1$. This is important since this choice will allow us to completely probe the high curvature region of the effective geometries.  

They can be obtained as in \cite{us-imp}. One gets
\begin{eqnarray}\nonumber
{}^{(0)}  g_{tt}(x) &=& -\left(1-\frac{r_S}{|x|+x_0}\right)\,,\\
{}^{(0)}  g_{tx}(x) &=& -{\rm sign}(x)\sqrt{\frac{r_S}{|x|+x_0}}\left(1+\frac{\delta x}{2(|x|+x_0)}\right)\,,\\\nonumber
{}^{(0)}  g_{xx}(x) &=& \left(1+\frac{\delta x}{2(|x|+x_0)}\right)^2, \quad {}^{(0)} g_{\theta\theta}(x)=(|x|+x_0)^2,\quad {}^{(0)} g_{\phi\phi}(x)=(|x|+x_0)^2\sin^2\theta.\label{eq:hatgmunu3}
\end{eqnarray}

For this slicing, the low curvature regions occur when $F(x)\simeq 0$ or equivalently at $x\to\pm\infty$. Concretely, at $x\to+\infty$ the effective metric approaches sufficiently fast a classical black hole metric in ingoing Painlev\'e-Gullstrand coordinates, while for $x\to-\infty$ the effective metric approaches sufficiently fast a classical white hole metric in outgoing Painlev\'e-Gullstrand coordinates. On the other hand, as we will see below, the curvature reaches its maximum when $F(x)= 1$, namely, at $x=0$.
\begin{figure}[ht]
{\centering     
  \includegraphics[width = 0.55\textwidth]{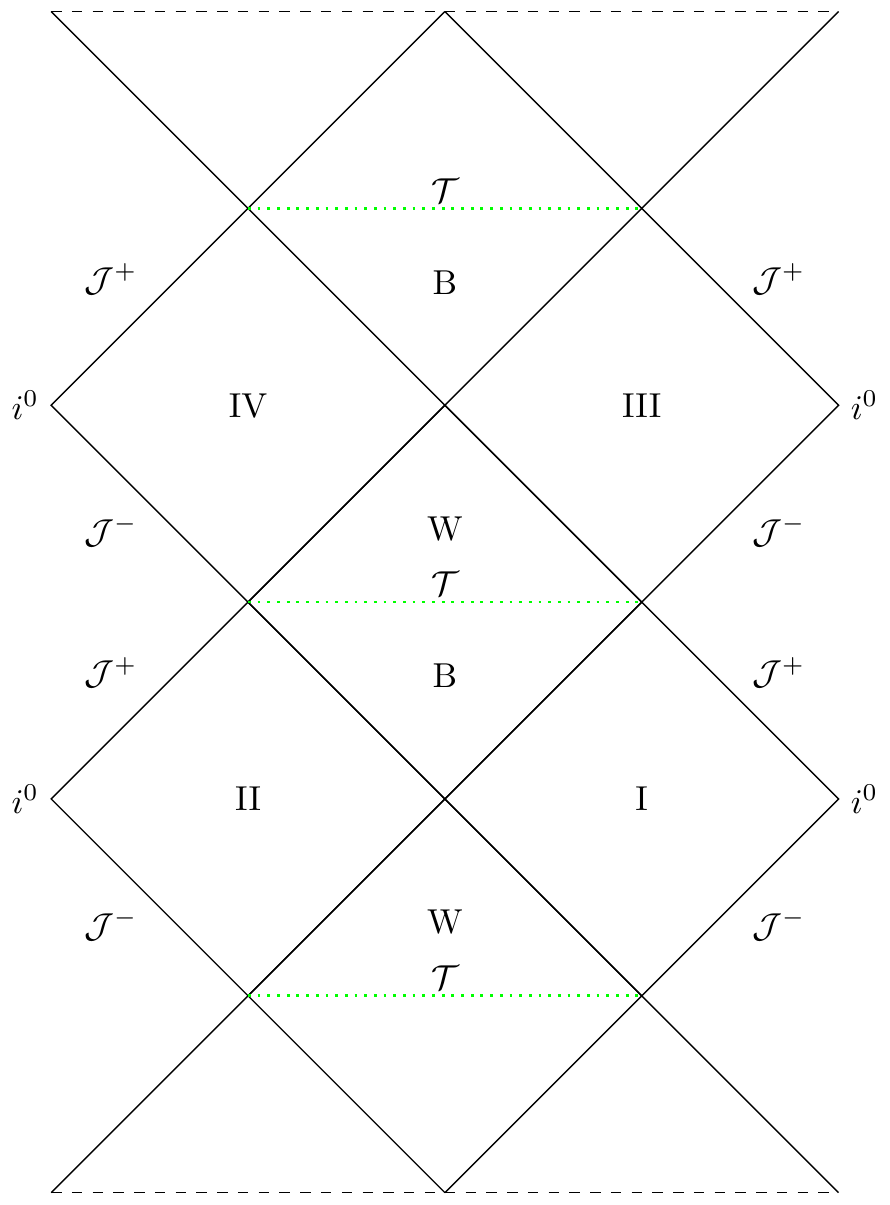}
}
 \caption{Penrose diagram of the effective geometry determined by the slicing in Eq. \eqref{eq:gf-f1}. 
 Black and green lines indicate low and high curvature regions, respectively. Continuous lines represent smooth regions while dotted lines are associated to a discrete geometry. Dashed lines indicate that the spacetime diagram continues up and down.  
 }
\label{fig:bh-wh}
\end{figure}\begin{figure}[ht]
{\centering     
  \includegraphics[width = 0.79\textwidth]{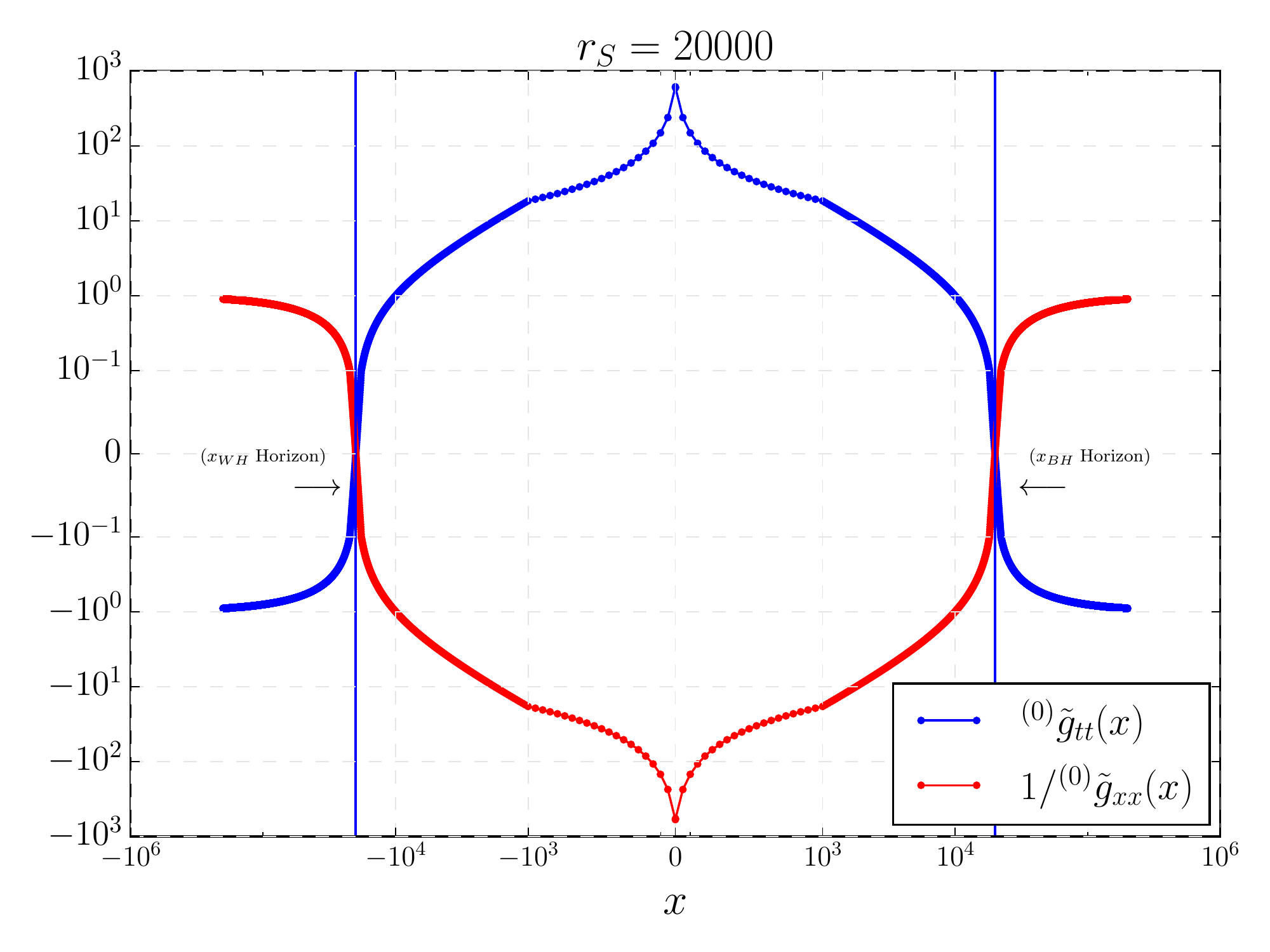}
}
\caption{The values of the $tt$ component of the metric and the inverse of $xx$ for the metric in diagonal form. When the first vanishes, horizons arise. Notice that in the region between the two horizons the discreteness is significant as represented in the separation of the dots (although in the plot we do not show all the points in the lattice but only one out of fifty).}
\label{gtt}
\end{figure}

In what follows, we refer to figure \ref{fig:bh-wh} {(see the similarities with the Penrose diagram of Ref. \cite{aos})}.
One can see that the condition ${}^{(0)}  g_{tt}(x)=0$ has two real solutions in $x$, corresponding to two classical black or white hole horizons, at $x_{BH}>0$ and $x_{WH}<0$. In the spacetime regions with $x>x_{BH}$ or $x<x_{WH}$, the surfaces $x={\rm const}$ are time-like, and correspond to untrapped regions. In the region right behind the black hole horizon, $x<x_{BH}$, $x={\rm const}$ hypersurfaces are space-like. This region is a trapped black hole interior. As we move towards the high curvature region, curvature is maximum at $x=0$. This space-like hypersurface connects the trapped black hole region with an anti-trapped white hole region. This is the so-called transition surface \cite{aos}. The anti-trapped white hole region extends all the way from $x=0$ to the white hole horizon $x=x_{WH}$. In all this region, $x={\rm const}$ hypersurfaces are still space-like. Once the white hole horizon $x=x_{WH}$ is crossed to the outside region, spacetime is untrapped again and $x={\rm const}$ hypersurfaces are again time-like.  

In order to illustrate all these properties, it is convenient to first write the effective metric in its diagonal form. (It should be noted that although the theory does not recover the full diffeomorphism invariance of the classical theory in the quantum regions, it is a valid mathematical tool to diagonalize a metric nevertheless.) It can be easily obtained by introducing the change of coordinates
\begin{equation}
d t \to dt+\frac{{}^{(0)}g_{tx}(x)}{{}^{(0)}g_{tt}(x)}dx
\end{equation}
This transformation amounts to the change 
\begin{equation}
{}^{(0)}g_{xx}(x) \to {}^{(0)}\tilde g_{xx}(x) = \frac{\left(1+\frac{\delta x}{2(|x|+x_0)}\right)^2}{\left(1-\frac{r_S}{|x|+x_0}\right)}, \quad {}^{(0)} g_{tx}(x) \to {}^{(0)}\tilde g_{tx}(x) = 0,
\end{equation}
while all other components remain as
\begin{align}
{}^{(0)}g_{tt}(x) &\to {}^{(0)}\tilde g_{tt}(x) = -\left(1-\frac{r_S}{|x|+x_0}\right), \\
{}^{(0)} g_{\theta\theta}(x) &\to {}^{(0)}\tilde g_{\theta\theta}(x) = (|x|+x_0)^2, \quad {}^{(0)} g_{\varphi\varphi}(x) \to {}^{(0)}\tilde g_{\varphi\varphi}(x) = (|x|+x_0)^2\sin\theta.
\end{align}

In figure \ref{gtt} we show two components of the effective metric in its diagonal form. There where they vanish, a horizon forms and the coordinate system becomes singular. However, we should remember that around $x\simeq x_0$ spacetime is discrete and the continuous line is just an interpolation. Therefore, the metric will be well defined provided the horizons are not located on a vertex of the lattice.

We have also studied the effective stress-energy tensor that encodes the main deviations from the classical theory. It is defined as
\begin{equation}
T_{\mu\nu}:=\frac{1}{8\pi G} G_{\mu\nu},
\end{equation}
where $G_{\mu\nu}$ is the Einstein tensor. $T_{\mu\nu}$ is characterized by the effective energy density $\rho$ and radial and tangential pressures densities, $p_x$ and $p_{||}$, respectively. They are defined by means of
\begin{equation}
\rho^{ext} := T_{\mu\nu}\frac{X^\mu X^\nu}{X^\rho X_\rho},
\end{equation}
\begin{equation}
p_x^{ext} := T_{\mu\nu}\frac{r^\mu r^\nu}{r^\rho r_\rho},
\end{equation}
and 
\begin{equation}
p_{||}^{ext} := T_{\mu\nu}\frac{\theta^\mu \theta^\nu}{\theta^\rho \theta_\rho},
\end{equation}
where $X^\mu$ is the Killing vector field that is time-like in the regions in which $x={\rm const}$ hypersurfaces are time-like. $r^\mu$ and $\theta^\mu$ are the vector fields pointing in the radial and $\theta$-angular directions, respectively. {When the Killing vector field $X^\mu$ is space-like, namely, in the regions in which $x={\rm const}$ hypersurfaces are space-like, $r^\mu$ becomes time-like. Therefore, 
\begin{equation}
\rho^{int} := T_{\mu\nu}\frac{r^\mu r^\nu}{r^\rho r_\rho},
\end{equation}
\begin{equation}
p_x^{int} := T_{\mu\nu}\frac{X^\mu X^\nu}{X^\rho X_\rho},
\end{equation}
while $p_{||}^{int} = p_{||}^{ext}$ since $\theta^\mu$ remains space-like. We will assume that these effective space-times can be approximated by a smooth and continuous geometry everywhere, even at the transition surface. This assumption, as we mentioned, fails in the most quantum region. However, we expect that $T_\mu^\nu$ (a quantity only valid when geometry is smooth) will still give us qualitative hints about quantum geometry corrections there.} 

\begin{figure}[ht]
  {\centering
  \includegraphics[width = 0.79\textwidth]{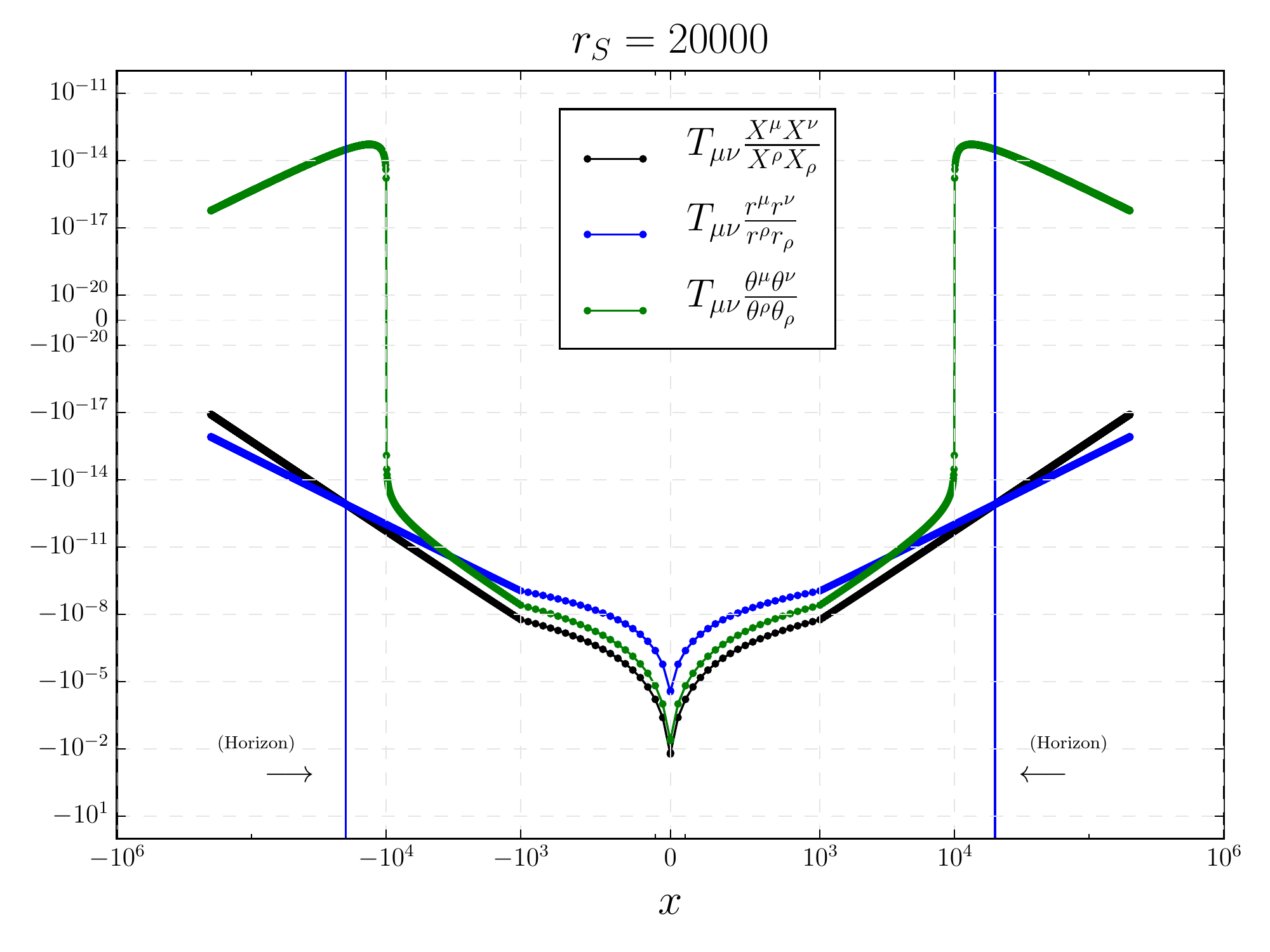}
}
\caption{The stress energy tensor of the effective metric ${}^{(0)}\tilde g_{\mu\nu}(x)$. This plot corresponds to $\delta x=\ell_{\rm Pl}$, namely, $s=1$.}
\label{tmunu}
\end{figure}

In figure \ref{tmunu} we show the components of the stress-energy tensor $T_{\mu\nu}$, or equivalently, the components of the Einstein tensor (up to a factor $(8\pi G)$)  {for the choice $\delta x=\ell_{\rm Pl}$. From them it is easy to extract the energy densities and pressures in each region of these effective space-times. } 

It is straightforward to compute the value of the energy density and pressures of the stress-energy tensor at the transition surface and in the limit of large mass $r_S\gg \ell_{\rm Pl}$. Actually, their value depend on the choice of spacing $\delta x$ of the uniform lattice in the radial direction. For instance, for $\delta x= x_0\left(\frac{\ell_{\rm Pl}}{x_0}\right)^s$ with $s=0,1,2$, one can see that\footnote{The choices of $\delta x$ shown here correspond to the maximum allowed uniform discretization if $s=0$, while $s=2$ gives the finest uniform refinement. $s=1$ is an intermediate choice.} 
\begin{eqnarray}\nonumber
\rho^{int}(x=0) &=& \frac{2\pi}{\Delta}\times{\cal O}\left(\left[\frac{\Delta}{2\pi r_S}\right]^{s/3+2/3}\right),\\\nonumber
p^{int}_x(x=0) &=& -\frac{2\pi}{\Delta}\times{\cal O}\left(\left[\frac{\Delta}{2\pi r_S}\right]^{s/3}\right), \\
p^{int}_{||}(x=0) &=& -\frac{2\pi}{\Delta}\times{\cal O}\left(\left[\frac{\Delta}{2\pi r_S}\right]^{s/3}\right)
\end{eqnarray}
Let us note that in the most quantum region, 
\begin{equation}
\omega_{x}(x=0)=\frac{p_x^{int}(x=0)}{\rho^{int}(x=0)} = -{\cal O}\left(\left[\frac{2\pi r_S}{\Delta}\right]^{2/3}\right),\quad \omega_{||}(x=0)=\frac{p_{||}^{int}(x=0)}{\rho^{int}(x=0)} = -{\cal O}\left( \left[\frac{2\pi r_S}{\Delta}\right]^{2/3}\right).
\end{equation}

As we see, at the transition surface, the effective stress-energy tensor does not violate the strong energy condition since $\rho^{int}(x=0)\geq 0$. However, it does actually violate the dominant energy condition. Since the dominant energy condition implies that $|\omega_x|\leq 1$ and $|\omega_{||}|\leq 1$, we conclude that this condition is violated since both $|\omega_{x}(x=0)|$ and $|\omega_{||}(x=0)|$ at the transition quantum spacetime blow up in the limit $r_S\gg \ell_{\rm Pl}$. 

One can construct the Penrose diagram of this geometry, together with a possible extension to regions not covered by our slicing.

\section{Discussion}

There are several comments about the scenario studied in this manuscript. On the one hand, the effective geometries that one can derive in this theory are uniquely determined by the semiclassical physical state and the (parameterized) observables that represent the components of the metric. {The quantum corrections on these geometries likewise depend on the minimal area gap $\Delta$ and the size of the discretization of the physical states we are considering. Polymer corrections due to the choice of foliation will also contribute if fluctuations of the mass are considered.} We are taking for simplicity an element (spin network) of the basis in the physical space of states and ignoring superpositions in different discretizations and masses. Quantum corrections break the covariance, in particular because their dependence on the discretization of the chosen quantum states, but also due to foliation dependent terms. The latter produce {$O(\Delta r_S^2/x^2)$} quantum corrections in the (asymptotically flat) external region of the black hole and therefore they are completely unobservable for macroscopic black holes, allowing to recover diffeomorphism invariance. Since different foliations are identified with (observer's) frames of reference, this is equivalent to say that, for physically implementable frames of reference (i.e. physically realizable observers) in the exterior region, quantum corrections will be negligible. Nevertheless, these quantum corrections increase when approaching the high curvature region, reaching maximum values of order {$O(\Delta r_S^2/x_0^2)$.} For instance, a free-falling observer (as it is the case under consideration in this manuscript) and {an accelerated observer will observe there only slightly different corrections, even if its foliation involves accelerations that are Planck order.}

{Regarding the original choice of shift as parametrized observable adopted in Ref. \cite{us-imp}, we noticed that, as mentioned in \cite{ewe-bh}, the most quantum region showed an inner Cauchy horizon connecting the trapped black hole region with a Planckian size transition space-time where $x={\rm const}$ hypersurfaces are time-like. However, strictly speaking, due to this Cauchy horizon, the extension beyond this region is not unique. After the bounce a Cauchy Horizon is traversed and therefore the initial conditions at ${\mathcal I}_-$ that end up producing a black hole are not enough for the determination of the possible extensions beyond the Cauchy horizon. Notice that the Cauchy horizon occurs in a deep quantum region that is in the past of the extension; further non-uniqueness would occur when quantum superpositions are considered. Besides, different foliations capture different extensions. We saw that a choice of foliation {(corresponding to an accelerated observer with a Planck order acceleration)} leads to an anti de Sitter universe beyond the Cauchy horizon. Similar ambiguities have been noted in classical general relativity \cite{dafermos}. One must keep in mind that these ambiguities can be alleviated by considering parametrized observables that correspond to physically implementable frames of reference (i.e. physically realizable observers). We are considering here extrinsic framings corresponding to a choice of polimerization for the functional parameter $K_\varphi(x_j)$. Even though the theory is covariant in the sense that the classical observables become quantum observables in the quantization process, each polimerization corresponds to a different choice of framing. In reference \cite{Gambini:2008ea} we proved that diffeomorphism invariance of the parametrized observables corresponding to the metric is only preserved for diffeomorphism that do not amplify Planck scale separation to macroscopic scale. The introduction of more realistic intrinsic framings resulting from the inclusion of matter would provide a natural choice of slicing allowing to solve this limitation.  For instance, the case of Painlev\'e-Gullstrand coordinates, that amount to a unit parametrized observable related to the lapse function. 
The kind of midisuperspace model here considered allows to analyze this issues while most of the minisuperspace scenarios proposed in the literature (see \cite{ashtekarsingh,bv,cgp, oss, cctr, bmm, sg-qd, qrlg-bh, adl} for references on hypersurface orthogonal slicings) adopted a particular family of space-time foliations where this issue of slicing dependence did not arise. Other authors have taken the issue of non-covariance to imply that modifications of the constraint algebra are in order, leading to the deformed hypersurface deformation algebra approach \cite{DHDA}.

Summarizing, we have applied an improved quantization scheme for loop quantum gravity in spherical symmetry. The singularity that appears in classical general relativity is eliminated and space-time is continued to a white hole space-time geometry through a transition surface where curvature reaches its maximum value. This is qualitatively similar to scenarios that have been recently proposed \cite{aos}. Our proposal yields effective geometries that are free of undesirable slicing dependencies in the semiclassical limit. Actually, the slicing independence in a precise semiclassical limit of small mass fluctuations can be invoked to restrict polymer modifications of the scalar constraint and the parametrized observables describing the quantum geometry. Finally, it is interesting to note that most of the ideas presented here and in Ref. \cite{us-imp} can be very useful in other situations, like in the vacuum polarized $T^3$ Gowdy cosmologies with local rotational symmetry \cite{gowdy}. }

\section{Acknowledgements}

This work was supported in part by Grant NSF-PHY-1903799, funds of the Hearne Institute for Theoretical Physics, CCT-LSU,  Pedeciba, Fondo Clemente Estable FCE\_1\_2019\_1\_155865 and Project.  No.  FIS2017-86497-C2-2-P of MICINN from Spain.  J.O. acknowledges the Operative Program FEDER2014-2020 and the Consejer\'ia de Econom\'ia y Conocimiento de la Junta de Andaluc\'ia.

\end{document}